\documentclass[12pt,a4paper]{article}
\usepackage{graphics}
\usepackage{cite,a4wide}
\newcommand{\beq}{\begin{equation}}
\newcommand{\eeq}{\end{equation}}
\newcommand{\bea}{\begin{eqnarray}}
\newcommand{\eea}{\end{eqnarray}}

\renewcommand{\a}{\alpha}
\renewcommand{\ni}{\noindent}

\newcommand{\g}{\gamma}
\renewcommand{\r}{\rho}

\newcommand{\s}{\sigma}

\newcommand{\E}{{\cal E}}

\newcommand{\oh}{\frac{1}{2}}

\newcommand{\non}{\nonumber}

\newcommand{\rf}[1]{(\ref{#1})}
\newcommand{\ra}{\rightarrow}

\begin{document}
\title{ Remarks on the Heavy Quark Potential in the
Supergravity Approach}

\author{\\ J. Greensite and P. Olesen \\
\\ The Niels Bohr Institute\\
Blegdamsvej 17\\
DK-2100 Copenhagen {\O}\\
DENMARK}
\maketitle

\bigskip
\bigskip

\begin{abstract}

   We point out certain unexpected features of the planar 
$QCD_3$ confining potential, as computed from a classical worldsheet action
in an AdS metric via the Maldacena conjecture.  We show that there
is no L\"{u}scher $c/R$ term in the static-quark potential, which is
contrary to both the prediction of various effective string models, and
the results of some recent lattice Monte Carlo studies.  It is also noted 
that the glueball masses extracted from classical supergravity tend 
to finite, coupling-independent constants in the strong coupling limit, even 
as the string tension tends to infinity in the same limit; this is
a counter-intuitive result.

\end{abstract}
\clearpage

    The startling possibility that continuum QCD may be solved 
(albeit in the large-N and strong-coupling limits) via classical 
supergravity has motivated a great deal of effort in recent months.  
Following the seminal
proposal of Maldacena \cite{Mal1,Mal2} and further developments by Witten 
\cite{Wit1,Wit2}, there have been explicit calculations of the string
tension \cite{Brand,Rey} and glueball mass spectrum \cite{Oog,Jev}
of non-supersymmetric planar QCD in D=3 and D=4 dimensions from the
supergravity approach.  A number of qualitative
features of planar QCD have been derived, such as the 
confinement/deconfinement transition 
at finite temperature \cite{Wit2}, screening of magnetic charge and
representation-dependence of Wilson loops \cite{Gross}, and the
length-law falloff of 't Hooft loops \cite{Li}, and these
agree with our expectations.  The glueball
mass ratios, although computed at strong couplings, have even been favorably
compared with corresponding lattice Monte Carlo results, extrapolated
to the continuum and large-N by Teper \cite{Teper}.  The supergravity
solution of planar QCD thus appears to be an explicit realization of
the large-N master field.

   In this article we will be concerned with certain qualitative
features of the heavy quark potential in the supergravity approach.
It should be noted that in computing the heavy quark potential there
are two types of contributions: those that come from the non-trivial
background metric (the ``classical'' worldsheet), and those that come
from quantum fluctuations of the worldsheet.  In the present paper shall
concentrate on the former, although at the end of the paper we will
also comment on the latter.

   Our main point is that there are two features of the heavy quark
potential in planar $QCD_3$, extracted along the lines of refs.\ 
\cite{Wit2,Brand,Rey}, which do not agree with the expected behavior
of the continuum theory.  These are the absence of a L\"{u}scher
term \cite{Luscher} in the potential, and the fact that the glueball
mass spectrum is almost independent of the string tension. 
We hasten to add that these points do not necessarily undermine the validity of
the Maldacena conjecture.  It only means that there
is some aspect of either the large-N or strong-coupling limits, as
taken in the supergravity approach, which causes the solution to differ 
qualitatively from finite-N, asymptotically free gauge theory. 

   We begin by calculating, in the supergravity approach, the subleading 
term in the static quark potential at large quark separation  
As explained in refs.\ \cite{Mal2,Wit2,Brand,Rey}, the heavy quark
potential in planar $QCD_3$ at strong-coupling is obtained by calculating
the action of a spacelike classical Nambu-Goto string in a Wick-rotated 
$AdS_5 \times S_5$ black-hole metric \footnote{We use throughout the notation 
of ref. \cite{Brand}.}
\beq
       ds^2 = \a' \left\{ {U^2 \over R^2}[f(U)dt^2 + dx_i^2]
                 + {R^2\over U^2} f^{-1}(U) dU^2 + R^2 d\Omega_5^2 \right\}
\eeq
where the boundary $C$ of the string worldsheet lies at $U=\infty$, and
\beq
      f(U) = 1 - {U_T^4 \over U^4},  ~~~~~~  
      R^2 = \sqrt{4\pi g_s N} = \sqrt{4\pi g_{YM}^2 N}
\eeq
It will also be useful below to make the rescaling
\beq
       U = \sqrt{4\pi g_s N} \rho ~,~~~   U_T = \sqrt{4\pi g_s N} b
\eeq
and express the metric in the form
\beq
       ds^2 = \a' \sqrt{4\pi g_s N} \left\{ {d\r^2 \over \r^2 - 
{b^4\over \r^2} } + (\r^2 - {b^4\over \r^2}) dt^2 + \r^2 dx_i^2 
                           +  d\Omega_5^2 \right\} 
\eeq
At length scales much less that $b^{-1}$, the supergravity solution is
probing planar ${\cal N}=4$ super Yang-Mills theory in D=4 dimensions,
while at scales above $b^{-1}$, the solution should probe planar
non-supersymmetric Yang-Mills theory in D=3 dimensions. The mass
scale $b$ can therefore be thought of as an ultraviolet regulator for 
the D=3 dimensional theory.  The proposal of refs.\ \cite{Mal2,Wit2,Brand,Rey}
is that the expectation value of the Wilson loop
at large N is given by $W(C) \sim \exp(-S)$, where the exponent is the 
action of the classical worldsheet bounding loop $C$.  
This leads to the following implicit expressions for the static quark
potential $E$ and a function of quark separation $L$
\begin{equation}
L=2\frac{R^2}{U_0}\int_1^\infty \frac{dy}{\sqrt{(y^4-1)(y^4-1+\epsilon)}},
\label{l}
\end{equation}
and
\begin{eqnarray}
E&=&\frac{U_0}{\pi}\int_1^\infty dy\left(\frac{y^4}{\sqrt{(y^4-1)(y^4-1+
\epsilon)}}-1\right)+\frac{U_0-U_T}{\pi}\nonumber \\
&=&\frac{U_0^2}{2\pi R^2}L+\frac{U_0}{\pi}\int_1^\infty dy\left(\sqrt{
\frac{y^4-1}{y^4-1+\epsilon}}-1\right)+\frac{U_0-U_T}{\pi}.
\label{e}
\end{eqnarray}
Here
\begin{equation}
\epsilon=f(U_0) \ll 1,~{\rm so}~U_0\approx U_T=\pi R^2 T=\sqrt{4\pi^3g_sN}T.
\end{equation}
We now want to obtain the next to leading behavior of $E$. To this end,
consider the integral 
\begin{equation}
J(\epsilon)\equiv\int_1^\infty dy\left(\sqrt{
\frac{y^4-1}{y^4-1+\epsilon}}-1\right).
\end{equation}
Clearly $J(0)=0$. In order to obtain an asymptotic expansion for small
$\epsilon$, it turns out to be most convenient first to differentiate 
$J(\epsilon)$,
\begin{equation}
\frac{\partial J(\epsilon)}{\partial\epsilon}=-\frac{1}{2}\int_1^\infty
\frac{\sqrt{y-1}}{(y-1+\epsilon/4)^{3/2}}\Phi (y),
\end{equation}
where $\Phi(y)$ is given by
\begin{equation}
\Phi(y)\approx\frac{\sqrt{(y+1)(y-i)(y+i)}}{(y+1-\epsilon/4)^{3/2}
(y-i+i\epsilon/4)^{3/2}(y+i-i\epsilon/4)^{3/2}},
\end{equation} 
which is regular for $y=+1$ and/or $\epsilon=0$. To obtain the asymptotic 
expansion, we perform a partial integration,
\begin{eqnarray}
\frac{\partial J(\epsilon)}{\partial\epsilon}&=&-\frac{1}{2}\left[\left\{
-2\sqrt{\frac{y-1}{y-1+\epsilon/4}}+2\ln (\sqrt{y-1}+\sqrt{y-1+\epsilon/4})
\right\}\Phi(y)\right]_1^\infty \nonumber \\
&+&\int_1^\infty dy \left(-\sqrt{\frac{y-1}{y-1+\epsilon/4}}
+\ln (\sqrt{y-1}+\sqrt{y-1+\epsilon/4})
\right)\Phi'(y).
\end{eqnarray}
Thus we have
\begin{equation} 
\frac{\partial J(\epsilon)}{\partial\epsilon}=\frac{1}{8}\ln\epsilon+
\int_1^\infty dy \left(-\sqrt{\frac{y-1}{y-1+\epsilon/4}}
+\ln (\sqrt{y-1}+\sqrt{y-1+\epsilon/4})
\right)\Phi'(y)+O(\epsilon).
\end{equation}
Making a further partial integration in the integral on the right hand
side of this equation, using
\begin{eqnarray}
&&\int dy\left(-\sqrt{\frac{y-1}{y-1+\epsilon/4}}
+\ln (\sqrt{y-1}+\sqrt{y-1+\epsilon/4})\right)\nonumber \\
&=&-\frac{3}{2}\sqrt{(y-1)
(y-1+\epsilon/4)}+(y-1+\frac{3\epsilon}{8})\ln(\sqrt{y-1}+\sqrt{y-1
+\epsilon/4}),
\end{eqnarray}
we see that
\begin{equation}
\int_1^\infty dy \left(-\sqrt{\frac{y-1}{y-1+\epsilon/4}}
+\ln (\sqrt{y-1}+\sqrt{y-1+\epsilon/4})
\right)\Phi'(y)=O(\epsilon\ln\epsilon).
\end{equation}
Collecting our results, we thus obtain
\begin{equation}
\frac{\partial J(\epsilon)}{\partial\epsilon}=\frac{1}{8}\ln\epsilon+
O(\epsilon\ln\epsilon).
\end{equation}
Since $J(0)=0$, we can integrate to obtain
\begin{equation}
J(\epsilon)=\frac{1}{8}\epsilon\ln\epsilon+O(\epsilon^2\ln\epsilon).
\end{equation}
In order to get a physical interpretation of this, we need to express $L$ in
terms of $\epsilon$. From eq. (\ref{l}) we have
\begin{equation}
L\approx 2\frac{R^2}{U_T}\int_1^\infty\frac{dy}{\sqrt{(y-1)(y-1+\epsilon/4)}}
\frac{1}{\sqrt{F(y)}},
\label{sing}
\end{equation}
where the function $F(y)$ is given by
\begin{equation}
F(y)=(y+1)(y-i)(y+i)(y+1-\epsilon/4)(y-i+i\epsilon/4)(y+i-i\epsilon/4).
\end{equation}
This function does not vanish for $y=1$ and/or $\epsilon=0$. Hence it does
not produce any singularity in the integral in eq. (\ref{sing}). To obtain
an asymptotic expansion of $L$, we proceed as before by a partial integration 
in eq. (\ref{sing}), using
\begin{equation}
\int \frac{dy}{\sqrt{(y-1)(y-1+\epsilon/4)}}=2\ln (\sqrt{y-1}+\sqrt{y-1+
\epsilon/4}).
\end{equation}
We thus obtain to the leading order
\begin{equation}
L\approx-\frac{R^2}{2U_T}\ln\epsilon+O(\epsilon\ln\epsilon).
\end{equation}
Hence the energy becomes
\bea
E &\approx& \frac{U_T^2}{2\pi R^2}L~\left(1-\frac{1}{2}e^{-2U_TL/R^2}\right)
\non \\
  &\approx& {\sqrt{4\pi g_s N} \over 2\pi} b^2 L \Bigl( 1 - \oh 
                    e^{-2bL} \Bigr)
\label{result}
\eea
The leading correction to the linear potential is thus exponentially small,
for $Lb>1$, of order $L\exp (-2Lb)$.

For $QCD_4$ a similar calculation can be performed, again using the results of
\ \cite{Brand,Rey}. We shall not repeat the details, which are quite similar to
those reported above, but we just give the final result,
\begin{equation}
E\approx\sqrt{4\pi g_{YM}^2N}b^2L\left(1-\frac{1}{2}e^{-bL}\right).
\end{equation}
Again we see that the correction is exponentially small. 

It should be emphasized that this result is valid also if the leading
correction in $(4\pi g_{YM}^2N)^{-1/2}$ is included. The first non-leading
correction to the AdS$_5$ black hole metric was found in \ \cite{gub} to be
\begin{equation}
\frac{ds^2}{\alpha'\sqrt{4\pi g_sN}}=(1+\delta_1)\frac{d\rho^2}
{\rho^2-\frac{b^2}{\rho^2}}+(1+\delta_2)(\rho^2-\frac{b^4}{\rho^2})dt^2
+\rho^2dx_i^2+d\Omega_5^2, 
\end{equation}
where 
\begin{eqnarray}
\delta_1&=&-\frac{15}{8}\zeta (3)\alpha'^3\left(5\frac{b^4}{\rho^4}+
5\frac{b^8}{\rho^8}-3\frac{b^{12}}{\rho^{12}}\right),\nonumber \\
\delta_2&=&+\frac{15}{8}\zeta (3)\alpha'^3\left(5\frac{b^4}{\rho^4}+
5\frac{b^8}{\rho^8}-19\frac{b^{12}}{\rho^{12}}\right).
\end{eqnarray}
This will modify
the integrals in $L$ and $E$ by polynomials in 1/$y$, and they do not modify 
the crucial logarithmic singularity at $y=1$. In particular, the leading 
correction to the linear potential is again of the type 
$\s L\exp (-{\rm const.} \times L)$.
Since the higher order corrections are expected to be of the polynomial
type, there is not much chance of modifying the exponential approach to the 
linear potential.

In lattice QCD there have been many calculations of the
heavy quark potential for various gauge groups, and it is safe to say
that the linear asymptotic behavior is well established (see, e.g.,
the results of Bali et al.\ \cite{Bali} for the case of D=4 and SU(2)).
Current numerical evidence for the sub-leading L\"{u}scher term
$-c/L$ at large $L$ in the interaction potential
\begin{equation}
E=\sigma L-c/L + ...
\label{pot}
\end{equation}
is quite
convincing for the case of $Z_2$ lattice gauge theory in D=3 dimensions
\cite{Caselle}, but there are also strong indications of the existence
of this term in the most recent (and, for us, more relevant) data for D=3
lattice SU(2) gauge theory \cite{Teper}.   
The proposal (\ref{pot}) for the potential is inspired 
by string theories, where $c$ is proportional to the central charge. For 
superstrings, $c=0$, and hence the L\"{u}scher term is absent for such 
strings.  For bosonic strings we expect $c=(d-2)\pi/24$, if $E(L)$ is
the quark potential extracted from Wilson loops, or $c=(d-2)\pi/6$ if
$E(L)$ represents the mass of a flux tube created by a Wilson loop winding 
once through a periodic lattice of length $L$ in the flux-tube direction
\cite{Forcrand}.  The general conclusion is that for confining strings 
in lattice gauge theory, the constant $c$ agrees fairly well with the values 
obtained in \cite{Luscher} and \cite{Forcrand} for the bosonic string.

   Comparing \rf{pot} to our result (\ref{result}), we see that if the 
fits of the lattice Monte Carlo data are taken seriously, then there is a 
large discrepancy between
the actual QCD string as seen on a lattice at weak couplings, 
and the QCD string obtained from supergravity. 
This could indicate the existence of a phase transition obtained in the
supergravity approach as the effective Yang-Mills coupling $g^2_{YM} N$
is reduced.  This possibility has already been noted by Gross and
Ooguri \cite{Gross}, and it is significant that
such transitions are known to occur in lattice gauge theory. The
strong and weak coupling regimes on the lattice are separated by a 
roughening phase transition;
{\sl in the strong coupling phase there is no L\"uscher term, whereas this
term does exist in the weak coupling phase.} 

It should again be emphasized that the order by order corrections 
in $(4\pi g_{YM}^2N)^{-1/2}$ do
not improve the situation with respect to the discrepancy between 
the next to leading order potential. This is because the logarithmic
singularity at $y=1$ is not influenced by these polynomial corrections.
Of course, it is a possibility that if the corrections could be
computed non-perturbatively, then the situation might improve.

\begin{figure}[ht]
\centerline{\scalebox{.5}{\rotatebox{270}{\includegraphics{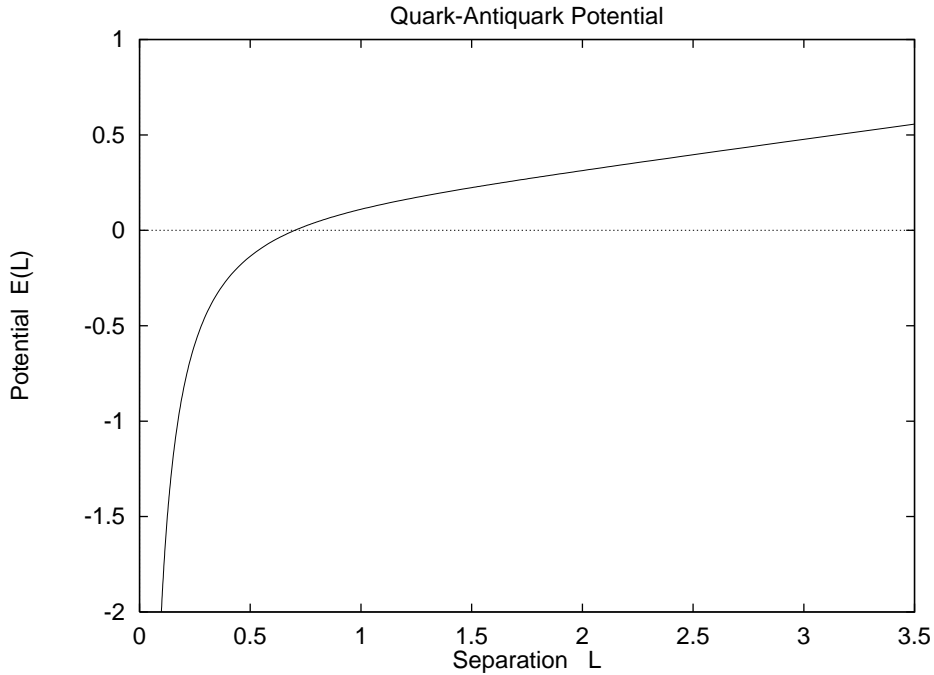}}}}
\caption{Heavy-quark potential determined from supergravity.  At short
distances $L<<1$, the potential is that of ${\cal N}=4$ super Yang-Mills
theory in D=4 dimensions, while for $L>1$ it should match that of
planar $QCD_3$ at strong coupling.}
\label{fig1}
\end{figure}

It goes without saying that there are always problems
extracting sub-leading behavior from a fit like
(\ref{pot}), since it is possible that a different type of fit may 
also reproduce the data quite well.  It is therefore interesting to ask 
what result would
be obtained if the same method for extracting $c$ on the lattice were
applied to potential derived from eqs.\ \rf{l} and \rf{e} above.  Put
another way, could the lattice results for $c$ be obtained from a potential of
the supergravity form? 
The method used in ref.\ \cite{Teper} was to compute the quantity
\beq
       c_{eff} = { {\E(\g L)\over \g L} - {\E(L)\over L} \over
                   {1 \over L^2} - {1\over (\g L)^2} }
\eeq
with $\g=1.5$, where in this case $\E(L)$ is the energy of a flux tube
state created by a spacelike Polyakov line, winding once through the 
periodic lattice.  The signal for the 
L\"{u}scher term is 
$c_{eff} \ra c$ as $L \ra \infty$, with $c$ finite.  In the numerical data 
presented in Table 9 of ref.\ \cite{Teper}, there is good evidence
of a systematic rise in $c_{eff}$ to a value consistent with
$c=\pi/6$, which is the ``universal'' (i.e. bosonic string) value appropriate 
to mass of flux loops on the periodic lattice.  

\begin{figure}[h]
\centerline{\scalebox{.5}{\rotatebox{270}{\includegraphics{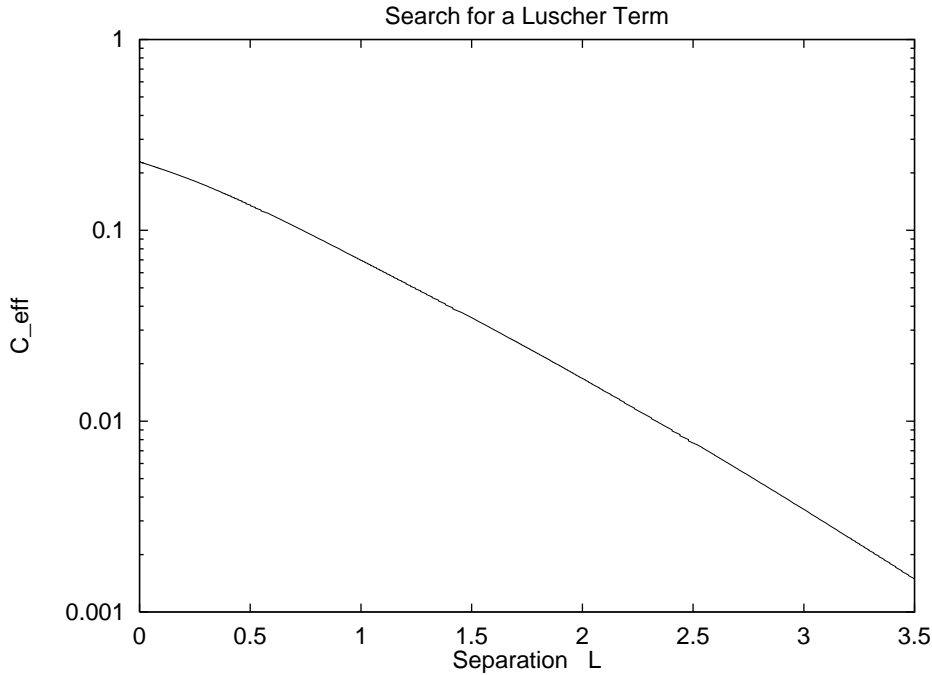}}}}
\caption{Searching for a L\"{u}scher term where there is no 
L\"{u}scher term: Calculation of $c_{eff}$ in the heavy-quark potential
obtained from supergravity.}
\label{fig2}
\end{figure}

   In Figure \ref{fig1} we show a numerical solution of eqs.\    
\rf{l} and \rf{e} for the static quark potential obtained from 
supergravity, where the axes display the rescaled, dimensionless values
\bea
        E_{rs} &=& {E(L) \over \sqrt{4\pi g_s N} b}
\non \\
        L_{rs} &=& Lb
\eea
In the region $L_{rs}\ll 1$ we are probing ${\cal N} = 4$ Yang-Mills in
D=4 dimensions, and the potential is Coulombic.  Hence $c_{eff}$ should
be constant in this region. The region of interest is $L_{rs}>1$, where
the solution probes planar QCD at strong-coupling in D=3 dimensions.
Figure \ref{fig2} is a plot of $c_{eff}$ vs.\ $L_{rs}$, extracted from
the potential shown in Fig.\ \ref{fig1}.  The point of this plot is
that $c_{eff}$ \emph{drops} with increasing $L$; precisely the opposite
behavior from what is reported in weak-coupling lattice gauge 
theory in D=3 dimensions \cite{Teper}.  Although this is hardly a conclusive
argument (it is always possible that at yet larger distances the 
Monte Carlo data for $c_{eff}$ will also start to drop), 
the existing evidence for flux-tube energy does
seem to favor $c \ne 0$ over a potential of the form derived from 
supergravity.   

   Finally, we comment briefly on the glueball mass spectrum of D=3
planar Yang-Mills, derived in the supergravity approach in refs.\ 
\cite{Oog,Jev}.  It is found that in the strong-coupling limit,
glueball masses have the form
\beq
        M_G = K\Bigl( 1 + O[(g^2_{YM}N)^{-3/2}] \Bigr) b
\eeq
where $K$ is a pure number which depends on the quantum numbers 
of the glueball in question, but which is \emph{independent of the
gauge coupling}.  Since the string tension obtained from \rf{result}
is
\beq
        \s = {\sqrt{4\pi g_s N} \over 2\pi} b^2
\eeq
the ratio 
\beq
        {M_G \over \sqrt{\s}} \ra 0
\eeq
tends to zero in the $g^2_{YM} N \ra \infty$ limit.      
This is a rather counter-intuitive result.  If glueballs are
thought of as tubes of electric flux, with the string tension essentially
the energy-per-unit-length of such flux tubes, then it is very natural
to expect the glueball mass to increase as the string tension increases.
This is certainly what happens in strong-coupling lattice theory.
But apparently it is not what happens in the supergravity approach.
We stress that this is a qualitative, rather than a numerical, issue; at 
strong-coupling there just seems to be no obvious relationship between 
the glueball mass spectrum and the strength of the confining force.

   We conclude that both of the features noted here, namely, the absence of 
a L\"{u}scher term in the heavy-quark potential, and the finiteness of the 
glueball mass in the limit of infinite string tension, suggest that
the supergravity solution to D=3 planar Yang-Mills 
theory is probably not a very realistic representation of continuum 
$QCD_3$.\footnote{Certain other reservations about the glueball spectrum
in the supergravity approach have been expressed in a very recent
article by Ooguri et al.\ \cite{Oog1}.}   
Of course, this solution was obtained in the strong-coupling limit.
It is therefore reasonable to expect that as the gauge-coupling is reduced, 
a phase transition of some kind is encountered, 
as also suggested in ref.\ \cite{Gross}.  Below this transition, which
is presumably associated with roughening, 
a finite L\"{u}scher term can appear in the potential, and a more 
realistic relation between string tension and glueball masses may be 
obtained.      

  In this article we have addressed only the heavy quark potential
extracted from the classical action, along the lines of refs.\
\cite{Mal2,Wit2,Brand,Rey}.  In principle there could be the possibility
that a L\"{u}scher term might arise in going beyond the classical action,
including also the quantum fluctuations of the worldsheet.  The worldsheet,
however, is that of a critical superstring.  Then $c=0$, at least naively, 
and one would not expect to get a L\"{u}scher term from this source.
However, there may still be a possibility, suggested to us by
Ooguri \cite{Oog2}, that worldsheet fluctuations in the neighborhood of
the horizon could produce an effective $c>0$.  Whether a L\"{u}scher
term of the appropriate magnitude could be produced in this way remains
to be seen.

\vspace{32pt}

\ni {\Large \bf Acknowledgements}

\bigskip

   We would like to thank Hirosi Ooguri for comments on an
earlier version of this article.
J.G.'s research is supported in part by Carlsbergfondet, and 
in part by the U.S.\ Department of Energy under Grant No.\ DE-FG03-92ER40711.


\begin{thebibliography}{xx}
\bibitem{Mal1} J. Maldecena, hep-th/9711200.
\bibitem{Mal2} J. Maldecena, Phys. Rev. Lett. 80 (1998) 4859, hep-th/9803002.
\bibitem{Wit1} E. Witten, hep-th/9802150.
\bibitem{Wit2} E. Witten, hep-th/9803131.
\bibitem{Brand} A. Brandhuber, N. Itzhaki, J. Sonnenschein, and
S. Yankielowicz, hep-th/9803263; hepth/98037137.
\bibitem{Rey} S-J. Rey and J-T. Yee, hep-th/9803001; 
S-J. Rey, S. Theisen, and J-T. Yee, hep-th/9803135.
\bibitem{Oog} C. Cs\'{a}ki, H. Ooguri, Y. Oz, and J. Terning, 
hep-th/9806021.
\bibitem{Jev} R. Koch, A. Jevicki, M. Mihailescu, and J. Nunes,
hep-th/9806125.
\bibitem{Gross} D. Gross and H. Ooguri, hep-th/9805129.
\bibitem{Li} M. Li, hep-th/9803252; hep-th/9804175.
\bibitem{Luscher} M. L\"{u}scher, Nucl. Phys. B190 [FS2] (1981) 317; \\
M. L\"{u}scher, K. Symanzik, and P. Weisz, Nucl. Phys. B173 (1980) 365.
\bibitem{Teper} M. Teper, hep-lat/9804008.
\bibitem{gub} S. S. Gubser, I. R. Klebanov, and A. A. Tseytlin,
hep-th/9805156
\bibitem{Bali} G. Bali, C. Schlichter, and K. Schilling, Phys. Rev. D51
(1995) 5165.
\bibitem{Caselle} M. Caselle, R. Fiore, F. Gliozzi, M. Hasenbusch, and 
P. Provero, Nucl. Phys. B486 (1997) 245, hep-lat/9609041.
\bibitem{Forcrand} Ph. de Forcrand, G. Schierholz, H. Schneider, and
M. Teper, Phys. Lett. B160 (1985) 137.
\bibitem{Oog1} H. Ooguri, H. Robins, and J. Tannenhauser, hep-th/9806171. 
\bibitem{Oog2} H. Ooguri, private communication.
\end{thebibliography}
\end{document}